\documentstyle[12pt]{article}

\newcommand{\be}{\begin{equation}}
\newcommand{\ee}{\end{equation}}
\newcommand{\ba}{\begin{eqnarray}}
\newcommand{\ea}{\end{eqnarray}}
\newcommand{\betafi}{\beta_{\Phi}}
\newcommand{\betasi}{\beta_{\sigma}}
\newcommand{\rozero}{\hat{\rho}(0)}
\newcommand{\rot}{\hat{\rho}(t)}
\newcommand{\tsig}{T_{\sigma}}
\newcommand{\msigma}{m_{\sigma}}
\begin{document}
\begin{flushright}
{PITT-93-4}
\end{flushright}
\begin{center}
{\bf DYNAMICS OF PHASE TRANSITIONS}\\
{\bf INDUCED BY A HEAT BATH}
\end{center}
\begin{center}
{{\bf Da-Shin Lee and Daniel Boyanovsky}}
\end{center}
\begin{center}
Department of Physics and Astronomy\\ University of
Pittsburgh \\
Pittsburgh, P. A. 15260, U.S.A.
\end{center}
\begin{abstract}
We study the non-equilibrium dynamics of a symmetry restoring
phase transition in a scalar field theory, the ``system'',
linearly coupled to another scalar field taken as
 a ``heat bath''. The  ``system'' is
 initially in an ordered low temperature phase, and the heat bath is
 at a temperature close to  the
critical temperature for the system. We estimate the time at which the
 phase transition to the disordered (symmetric) phase takes place.
 We derive, and integrate the one-loop effective equations of motion for
 the  order parameter  that include the effects of the heat bath.
 A semiclassical
Langevin equation is derived and it is found that it contains a
 non-dissipative, non-Markovian kernel, the noise term is colored
 and correlated on time scales determined by the temperature of
 the heat bath. The range of validity of the Langevin equation
  and a
 consistent procedure to incorporate corrections are discussed.
\end{abstract}

\section{\bf Introduction and Motivation}

Non-equilibrium aspects of quantum field theory and statistical mechanics
are beginning to receive considerable attention, for example
 within the context of structure formation in
inflationary cosmology\cite{kolb,linde,brandenberger,starobinsky},
dissipation and decoherence in quantum cosmology\cite{paz1,paz2}, in
 the theory of quantum brownian motion\cite{paz3,zurek,carlitz}, and field
theory in noisy environments\cite{cloutier} to cite but
a few.

An important setting in which statistical mechanics out of equilibrium
will play a very important role is in the description of the dynamics
of phase transitions in the early universe. Recently the effective
evolution equations to one loop order for the order parameter\cite{boyveg}
 and the process of domain formation and growth\cite{boyshin} during a
typical second order phase transition had been studied.

In this article we study the non-equilibrium aspects of a symmetry
restoring phase transition induced by coupling a system originally at low
temperature in a broken symmetry state to a heat bath at a temperature
close to the critical temperature of the system. Furthermore we
derive a Langevin equation for the system that incorporates the effects
of the heat bath, determine its range of validity, and provide a
 consistent procedure to incorporate corrections to the Langevin
description.

Phenomenological Langevin equations had been recently used in a variety
of field theoretical applications: for a semiclassical non-equilibrium
analysis of thermal activation\cite{gleiser1,bochkarev}, and the dynamics
of kink-pair production in a heat bath\cite{gleiser2}. These
 phenomenological Langevin equations incorporate a linear friction term
and an uncorrelated (white) noise that satisfy the classical limit of the
fluctuation-dissipation theorem.

Our motivations for this article are twofold. First we want to study the
non-equilibrium dynamics of a phase transition induced by coupling
a system  to a heat bath from a first-principle calculation. Secondly
we want to qualify and quantify the situations in which a phenomenological
Langevin equation with friction and white noise is justified, again,
by establishing a first-principle derivation of the field-theoretical
Langevin equation, its range of validity and possible corrections.
 One of the physically relevant questions that we want to
address is about the time scales in which a ``system''
equilibrates with a heat-bath.

An equilibrium description of a thermodynamic system assumes
that the degrees of freedom under consideration had reached
thermal equilibrium with a ``heat bath'' or reservoir, and that the
system is observed at times much larger than its typical relaxation time.
 When all the dynamical transient effects have had
enough time to relax, the final state of the system would be that of
equilibrium and the final state will not be sensitive to details of
initial conditions or correlations, but only to the global thermodynamic
properties of the ``heat bath'', like temperature etc.

Although these assumptions are commonly used and clearly
correct in a wide variety of experimental situations, their extrapolation
to the realm of a Quantum Field Theory in extreme environments like
the Early Universe or Heavy Ion Collisions is not only unclear but
perhaps unwarranted a-priori. In particular, the concept of a ``heat
bath'' is somewhat obscure when studying a field theory as a
``system''. Strictly speaking, a ``heat bath'' or reservoir, must have
 certain characteristics that allow to separate in a well defined manner
 the reservoir from the
system, in particular the characteristic relaxation times
of the heat bath must be {\it much shorter} than those of the system,
secondly, the reservoir must have an infinite specific heat (a condition
necessary for the equivalence between a microcanonical and canonical
description) and thus infinitely many degrees of freedom.

 The second condition is fulfilled if the heat bath
is a quantum field theory, but if the ``system'' under consideration is
{\it also} a quantum field theory, both system and bath have infinitely
many degrees of freedom and the separation becomes somewhat artificial.

 The separation of time scales presents another conceptual difficulty
 in field theory  because both system and
 bath  have degrees of freedom at all energy scales with typically
 comparable density of
states at large energies (determined solely by the
relativistic dispersion relation).

If the relaxation time for the heat bath is much shorter than that of the
system, one may ignore the dynamics of the heat-bath and only ask questions
about the system. Effectively this is achieved by ``tracing over'' the
degrees of freedom of the bath.

In this article we examine the situation in which a self
interacting scalar
field is coupled to another scalar field at  a particular
time. Prior to
this time both fields are uncoupled and assumed to be
described by a
thermal equilibrium ensemble, but at different temperatures.
 This
initial condition is somewhat arbitrary and unjustified, but
our purpose
is to study the time evolution of one of the fields (the
self-interacting
scalar field) taken to be the ``system'', and trace out the
other field,
taken to be the ``heat bath''. If the ``system''
field is
at zero temperature in a state of broken symmetry, how long
does it take
to produce the phase transition and to restore the symmetry
if the ``heat
bath'' is at a temperature close to the critical??.

A very succesful approach to study a Brownian particle in a heat bath or
environment is provided by the Langevin equation. The effect of the heat
bath is to introduce a friction term and a stochastic noise (typically
assumed to be uncorrelated) both being related by the
 fluctuation-dissipation theorem.

We want to study if such a description is suitable for a {\it field theory}
and if so under which circumstances, establishing the range of validity of
such description and possible corrections.

In section II the essential elements to study field theory out of
 equilibrium
are introduced. In section III we study the dynamics of the symmetry
restoration phase transition induced by the heat bath.  Section IV presents
the derivation of the Langevin equation, discusses the range of validity
and the formal procedure to incorporate corrections.
 Our conclusions are summarized
in section V. Two appendices present many technical details.

\section{\bf Evolution out of Equilibrium: Real Time
Analysis}

As mentioned in the introduction, our aim is to study the
time evolution
of the ``system'' when it is coupled to a  ``heat bath'' at
a different
temperature. By ``system'' we now refer to the simplified
case of a
self-interacting scalar field, and by ``bath'' we will take
for simplicity
the case of a free massive field with a  mass much larger
than the  mass of the
quanta of the ``system'', so as to justify that the time
scales of the heat bath are much shorter than those of the system.
Eventually we will take the temperature of the heat bath to be
much larger than the masses (of both system and bath). In this situation,
we can relax the constraint that the mass of the particles in the heat
bath be much larger than that of the system.

One can avoid having to face the question of how the ``bath''
acquired its temperature, by assuming that the bath degrees of
freedom are in a highly excited state corresponding to a microcanonical
ensemble with an energy density ${\cal{E}} \gg m_{\sigma}$ with
$m_{\sigma}$ the mass of the particles in the bath. Under these
circumstances one can pass from the microcanonical ensemble at
this energy density to a canonical ensemble by a Laplace transform.
 The temperature of the equivalent canonical ensemble is
 $T_{\sigma} \approx {\cal{E}}^{\frac{1}
{4}}$. We will use the  description in terms of the canonical ensemble.

We will also require that the coupling of the system to the
bath satisfies
the following conditions: i) locality ii) manifest Lorentz
covariance, iii) maintains the renormalizability  of
the scalar field theory.

For calculational simplicity we will also assume that the
coupling is linear in both fields.
 The simplification of a ``free field bath'' and linear
coupling will allow
us to present detailed calculations, these restrictions may
be relaxed at the expense of further complications.

Thus our situation may be modelled by introducing the time
dependent Hamiltonian

\be
   H(t) = H_{\Phi} + H_{\sigma} + H_I(t)
\ee
where

\be
  H_{\Phi}  =  \int_V   d^3x \left\{
\frac{1}{2}\Pi^2_{\Phi}(x)+
\frac{1}{2}(\vec{\nabla}\Phi(x))^2+\frac{1}{2}
(-\mu^2) \Phi^2(x) +\frac{\lambda}{4!}\Phi^4(x) \right\}
\ee
is the Hamiltonian of the self-interacting field that will
be taken as
the ``system'', and  $\mu^2$ is chosen to be positive with
the purpose of
studying non-equilibrium aspects of phase transitions.

\be
  H_{\sigma}  =  \int_V   d^3x \left\{
\frac{1}{2}\Pi^2 _{\sigma} (x)
+\frac{1}{2}(\vec{\nabla}\sigma(x))^2+\frac
{1}{2} m^2_{\sigma}  \sigma^2(x)  \right \}
\ee
is the Hamiltonian for the free field taken to be as
the ``bath''. Finally  $H_I(t)$
is the interaction Hamiltonian given by

\be
  H_I(t)  =   g \Theta(t)  \int_V  d^3x \Phi(x) \sigma(x)
 \label{interham}
\ee

To completely determine the time evolution of the ``system-
bath'', we
need to specify the initial conditions. We assume (without
any a-priori
justification), that for $t < 0$ both the system and bath
are  in
thermal equilibrium at temperatures $T_{\Phi} \; ; \;
T_{\sigma}$
 respectively. Furthermore we will assume that the field
$\Phi$ is in
a broken symmetry state at $T_{\Phi} \approx 0$ and that
$T_{\sigma}
\gg m^2_{\sigma}, \mu^2$.
Notice that the coupling constant $g$ has dimensions of $(mass)^2$, and
a perturbative expansion in this coupling, will involve the ratio of
this coupling to a particular energy scale of the heat bath.
Under the assumption that the temperature of the bath is much larger
than the masses of the bath and system field, we expect the perturbative
expansion parameter to actually be in terms of the dimensionless ratio
 $\tilde{g}=g/T_{\sigma}^2$.
Furthermore, in this limit, we expect the characteristic time (relaxation
time) of the heat-bath to be $t_{bath} \approx 1/T_{\sigma}$.

For the study of the dynamics in real time, the quantity of
interest is
the density matrix in the Schroedinger picture. For $t \leq
0$ the system and bath are uncoupled and  the initial density matrix
 is given by

\be
\hat{\rho}(0) = e^{-\betafi H_{\Phi}} \otimes e^{-\betasi H_{\sigma}}
\ee
where $\beta = 1/K_BT$ for the respective temperatures.
In the Schroedinger picture, the density operator evolves in
time as

\be
  \hat{\rho}(t) =  U(t) \hat{\rho}(0) U^{-1} (t)
 \ee
with $U(t)$ the time evolution operator.

As mentioned previously, our aim is to understand the
dynamics of a
field interacting with a heat bath. Physically we are
interested in
correlation functions of the scalar field $\Phi$, but not on
any properties of the heat bath.
 Thus, we proceed to trace over the degrees
of freedom
of the heat bath, obtaining an effective theory for the
scalar field.

This procedure, starting from a closed ``system-heat bath''
and tracing over
the reservoir degrees of freedom  leads to the
description of  the ``system''
as an open environment, and is fundamentally equivalent to
passing from a microcanonical to a canonical description.

The procedure of tracing out (performing the path integral) the
bath degrees of freedom yields to a ``reduced density matrix''
for the system. Although the full density matrix satisfies the
Liouville equation with the full Hamiltonian,
 the reduced density matrix  satisfies a
Fokker-Planck equation\cite{leggettcaldeira,pazhu}.

This procedure originally proposed by Feynman and
 Vernon\cite{feynmanvernon},
 has been used repeateadly in the literature\cite{leggettcaldeira},
 in particular
in the study of dissipative effects in quantum mechanics and
for the
consistent treatment of quantum Brownian motion\cite{grab}. There are
some major differences between our work and past work on quantum
brownian motion and dissipative systems.
In the usual treatment of quantum
brownian motion\cite{leggettcaldeira}, a particular density of states
 {\it must be assumed for the heat bath}.
The  choice of density of states leads to different
dissipative behavior,
and is reflected in the correlation functions of the heat
bath through the fluctuation-dissipation relation.

In our case, the nature of the heat-bath, and its coupling
to the system,
are constrained by the requirements of locality,
relativistic covariance, and renormalizability,
 which severely constrain the density of states
of the heat bath and the possible couplings to the system.

The expectation value of an arbitrary operator
 $\langle {\cal {O}}\rangle(t)$ is given by

\be
\langle {\cal{O}} \rangle(t) =\frac{ Tr \rot {\cal{O}}}{Tr
\rozero }
\label{average}
\ee

It is more illuminating and convenient for our purposes to
cast the
above expression in terms of the time evolution operator U.
This is
achieved by first choosing a large negative time $T <0$ (not
to be
confused with temperature), for which $U(T)= U_{\Phi} (T)
U_{\sigma} (T)$
 where $U_i(T)=\exp[-iH_i T]$
 (for $i=\Phi\; ; \; \sigma$) we may write
 $ \exp[-\beta_i H_i]=\exp[-iH_i(T-i\beta_i-T)]=U_i(T-
i\beta_i, T)$.

 Inserting in the trace  $U^{-1}(T) U(T)=1$,
 commuting $U^{-1}(T)$
 with $\hat{\rho}(0)$ and using the composition property of
the
 evolution operator, the expectation value (\ref{average})
becomes

\be
\langle {\cal{O}} \rangle(t) =\frac{ Tr \left\{ U_{\Phi}(T-
i\beta_{\Phi}, T)
 U_{\sigma} (T-i\beta_{\sigma}, T)
 U^{-1}(T,t)
{\cal{O}} U(t,T)   \right\} }{ Tr \left\{ U_{\Phi} (T-
i\beta_{\Phi}, T)
 U_{\sigma}  (T-i\beta_{\sigma}, T)   \right\} }
\label{average2}
\ee

It proves convenient to choose a large positive time $T'$
and
 to insert $U(t,T')U(T',t)=1$ (with U the {\it full time
evolution
 operator})  to the left of
 ${\cal{O}}$ in (\ref{average2}) to extend the contour in
the numerator
 to an arbitrary large positive time $T'$\cite{mills}.
  Finally, the thermal expectation value of the operator
 $ {\cal{O}}$ is

\be
\langle {\cal{O}} \rangle(t) =\frac{ Tr \left\{  [U_{\Phi} (T-
i\beta_{\Phi}, T) U_{\sigma}
 (T-i\beta_{\sigma}, T) ] U(T,T') U(T',t)
{\cal{O}} U(t,T)   \right\} }{ Tr \left\{ U_{\Phi} (T-
i\beta_{\Phi}, T) U_{\sigma}
 (T-i\beta_{\sigma}, T)   \right\} } \label{expectvalue}
\ee

In the  complex time plane, the numerator represents the
following
 process:
 from $T<0$, evolve in time up to $t$, insert   the operator
 ${\cal{O}}$,  evolve further up to $T'$, and backwards from
$T'$ to
 $T$, finally down the negative imaginary axis to $T-
i\beta_{\Phi}$ for
 $\Phi$ field ($T-i\beta_{\sigma}$ for the $\sigma$ field) depicted in
 figure (1).
 The denominator just evolves along the negative axis from
 $T$ to $T-i\beta_{\Phi  ; \sigma}$ for $\Phi \; ;
\sigma$.
 Eventually we will take the arbitrary times $T \rightarrow
-\infty
\; ; \; T' \rightarrow \infty$.

 The insertion of the operator ${\cal{O}}$ may be obtained
as usual
 by introducing external   sources  coupled to the
particular operator.

Since we propose to use perturbation theory to study the
non-equilibrium
correlation functions, it proves more convenient to
introduce different
 source
terms for {\it all} the time evolution operators in the
trace. The
 sources will be different on the different branches along
the complex
time contour. However, we do not introduce source terms for
the field
$\sigma$ as it will be traced out, and no correlation
functions of this field will be computed.
We are thus led to consider the following generating
functional

\be
Z[J^+_{\Phi}, J^-_{\Phi}, J^{\beta}_{\Phi}]= Tr\left\{ [U_{\Phi} (T-
i\beta_{\Phi},T;J^{\beta}_{\Phi}) U_{\sigma} (T-i\beta_{\sigma},T)]
U(T,T';J^-_{\Phi})U(T',T;J^+_{\Phi}) \right\} \label{partfunc}
\ee
where $J^{\pm}_{\Phi}$ are the
source terms for the field $\Phi$ along the forward ($+$)
and backward ($-$)
segments of the contour, and the denominator in
 (\ref{average2}) is given by $Z[0,0,J^{\beta}]$. This formalism allows
to study more general situations with different choices of $\rho(0)$.

The necessity for these generating functionals to study non-equilibrium
quantum statistical mechanics was originally proposed by
Keldysh\cite{keldysh} and Schwinger\cite{schwinger} and has been used
in the literature very

often\cite{landsman,semenoffweiss,jordan,niemisemenoff,kobeskowalski,calzetta,calzettahu}.
The trace over the  $\sigma$ field (functional integral) may
be easily
performed because it is a free field linearly coupled to the
scalar field
$\Phi$, which acts as a ``source term'' for the $\sigma$
field, yielding
to a  non-local quadratic {\it influence functional}\cite{feynmanvernon}.

\be
 {\cal{F}}[\Phi^+,\Phi^-] = \exp \left\{ - \frac{i} {2} g^2
\int^{T'}_0 dt_1
 \int^{T'}_0 dt_2
           \Phi^a(t_1) D^{ab}_{\sigma}(t_1,t_2)  \Phi^a(t_2)
\right\}\label{influfunc}
\ee
with a sum over the indices $ a,b = +,- $, corresponding to the forward
$T \rightarrow T'$ and
backward $T' \rightarrow T$ branches of the contour
respectively, and
$( D^{ab}_{\sigma}(t_1,t_2))$ are the Green's functions on
the contour for the ``heat-bath'' given in  appendix A.

We relegate  most of the technical details to  the appendices,
where we
show that the final form of the generating functional
becomes

\begin{eqnarray}
Z[J^+_{\Phi}, J^-_{\Phi}] & = &  \exp \left\{ i
\frac{\lambda}{4!}
 \int^{\infty}_{-\infty} dt  [ (-i \frac{\delta}{\delta
J^+_{\Phi}})^4
 - (i\frac{\delta}{\delta J^-_{\Phi}})^4 ] \right\}
{\cal{F}}[ -i
 \frac{\delta}{\delta J^+_{\Phi}}, i \frac{\delta}{\delta  J^-
_{\Phi}}]
 \times  \nonumber\\
        &   & \exp \left\{ - \frac{i} {2} \int^{\infty}_{-
\infty}  dt_1  \int^{\infty}_{-\infty} dt_2
           J^a_{\Phi} (t_1) D^{ab}_{\Phi}(t_1,t_2)  J_{\Phi}^a (
t_2) \right\}
 \label{generatingfunction}
\end{eqnarray}
with $a,b = +,-$.

The Green's functions on the contour and their properties
are analyzed in  appendix A.
We are now in condition to study the dynamics {\it in real time}
for the field $\Phi$ in the presence of the ``heat-bath''. As mentioned
in the introduction, one of our principal motivations is to address
the fundamental question of symmetry restoration by the heat bath. In
particular, if the system is originally at very low temperatures (or zero
temperature) in a broken symmetry state and is suddenly coupled (at $t=0$)
 to a ``heat bath'' in equilibrium at a temperature close to the
 critical temperature,
 how long does it take for  the symmetry to be restored?? What are the
relevant time scales for the dynamics of symmetry restoration?? The
answer to the last question is certainly far from obvious. There are
several widely different time scales in the problem, the typical
time of the heat bath (determined by the mass and temperature
 of the $\sigma$ field),
the mass and temperature of the scalar field (system),
  and finally the coupling
system-bath. It is not a-priori obvious which time scale or combination
thereof will determine the dynamics of the process of
 symmetry restoration.
However, on physical grounds, we expect that for very weak coupling
between the system and the heat bath, this time will be very large. This
intuitive argument must be qualified however. The dimensionless coupling
is expected to be $g/T^2_{\sigma}$, and the transition time is expected
to be ``large'' on time scales of the heat bath.  Thus when the
 temperature of the bath is much larger than the masses of both system
and bath we expect the ``transition time'' $t_c$ (the time at which the
symmetry restoration phase transition occurs) to be such that

\[ t_c T_{\sigma} \gg 1 \]

\section{\bf Symmetry Restoration}

Usually the physics of a phase transition is studied by means of the
 {\it static} effective potential (free energy density),
 as the position of the
 minima of the effective
potential determine the thermodynamic equilibrium states of the system. If the
minima
correspond to a nonzero value of the expectation value (or thermal
average) of the scalar field the symmetry is spontaneously broken,
otherwise the symmetry is restored. The vacuum expectation value or
 thermal ensemble  of the scalar field serves as the order parameter for
the transition. In a translational invariant theory this order parameter
is independent of the spatial coordinates, and {\it in equilibrium} it
must be {\it independent of time}.

In our case, we couple suddenly the system to the heat-bath at time
$t=0$ and let the system evolve in time with the full interaction
Hamiltonian. Clearly this is {\it not} an equilibrium situation, as
the initial density matrix does not commute with the full Hamiltonian
for $t>0$. The static effective potential is not the proper
quantity to study the {\it dynamics} of the phase transition and one
should invoke the effective action to account for the real time
dependence of the situation.  Because translational invariance
is still preserved, the order parameter
\be
\langle \Phi(\vec{x},t) \rangle = \frac{Tr\rot \Phi(\vec{x},t)}
{Tr \rozero} = \phi(t)  \label{orderparameter}
\ee
only depends on time (and certainly temperature).

Initially  the scalar field is in equilibrium in the
 broken symmetry phase at very low temperature ($T_{\Phi}
\ll T_c =  \sqrt{24\mu^2/\lambda}$) and is suddenly coupled to
the heat bath whose temperature is $ T_{\sigma} \approx T_c \gg
 m_{\sigma} \; , \; \mu \;$.
 The order parameter will evolve in time and the
phase transition will occur when the order parameter becomes zero or
begins to oscillate around zero.

Obtaining the effective action even to one-loop order is clearly an
imposing task,  the same information about the phase transitions
is obtained from the {\it effective evolution equations} for the
order parameter. These equations of motion are formally obtained as
a variational principle from the effective action, by requiring that
the functional derivative of the effective action with respect to the
order parameter vanishes. It turns out that it is much easier to obtain
the effective equations of motion for the order parameter from the
non-equilibrium formalism. This is achieved by using the tadpole
 method\cite{weinberg} and applied to this non-equilibrium situation
by shifting the fields $\Phi^{\pm}$ on the forward and
backward branches as follows\cite{boyveg,boyshin}(see appendix B)

\be
\Phi^\pm  (\vec{r},t) = \phi (t) + \varphi^{\pm}(\vec{r},t)
\ee
The reason for shifting both $(\pm)$
fields by the
 same  configuration, is that $\phi$ enters in
the time evolution
 operator as a background c-number field, and evolution
forward
 and backwards are now considered in this background.
The effective equations of motion are obtained by requiring that
\be
      \langle \varphi^{\pm}  (\vec{r},t)  \rangle  =  0
\ee
These two conditions consistently result in the {\it same}
evolution equation for the order parameter $\phi$ as a consequence of the
cyclic property of the trace.

 As explained in appendix B the evolution equations are obtained by
expanding the
 action in (\ref{effectiveaction}) around $\phi(t)$, the quadratic terms
for $\varphi ^{\pm}$ will define the propagators and the higher order
 terms will be treated in perturbation theory. To one loop order we
find

\begin{eqnarray}
 \frac{d^2}{dt^2} \phi(t) + (-\mu^2)  \phi(t) +
\frac{\lambda}{3!}
 \phi^3 (t) & - & \frac{g^2}{m_{\sigma}} \int^t_0 dt'
 \sin[m_{\sigma} (t-t')] \phi(t') \nonumber\\
 & + & \frac{\lambda}{2} \phi (t) \int \frac{d^3
k}{(2\pi)^3}
 \langle  \varphi^+_{\vec{k}} (t) \varphi^+_{-\vec{k}} (t) \rangle = 0
\label{equationofmotion}
\end{eqnarray}

The retarded kernel in the above expression is arising from the
influence functional, notice that it {\it does not} depend on the
temperature of the heat bath, this is a consequence of the fluctuation-
dissipation theorem. This kernel is non-Markovian (has memory), but
perhaps more importantly it is {\it non-dissipative}.

We will see later that this is the retarded kernel entering in the
Langevin equation that describes the effective
semiclassical evolution equations in the presence of the heat bath.

The one-loop term $ \langle  \varphi^+_{\vec{k}} (t)
 \varphi^+_{-\vec{k}} (t) \rangle $ is recognized
 as the spatial Fourier transform
of the equal time two-point correlation function of the fluctuations.
 This correlation
function is obtained by inverting the operator of quadratic fluctuations.

 The quadratic terms that define these propagators, however,
  depend on the time dependent order parameter $\phi(t)$, and
also receive a contribution from the influence functional (that mixes
$\varphi^{\pm}$). The identification of the ``free propagator''
is extremely difficult as one must find the inverse of the
{\it time dependent} quadratic operator for the fluctuations.
This involves summing the Dyson series for the
mixing term in the influence functional. Because of the complicated
time dependence we were unable to sum up the Dyson series and have
to content ourselves with a perturbative expansion for both the
time dependence of the order parameter and the contribution of the
influence functional. Thus we write

\ba
-\mu^2 + \frac{\lambda}{2}\phi^2(t) & = &  M^2
 + \frac{\lambda}{2}(\phi^2(t)-\phi^2(0)) \nonumber \\
                               M^2  & = &
 -\mu^2 + \frac{\lambda}{2}\phi^2(0) \label{massterm}
\ea
with $\phi^2(0) = 6 \mu^2 / \lambda$ being the minimum of the {\it tree
level} potential. We identify the first term on the right hand side
of (\ref{massterm}) as the mass of quanta in the broken symmetry
state. The second term will necessarily be of order $\lambda g^2$
or higher,
thus ${\cal{O}}(\lambda)$ smaller than the term arising from the
influence functional
because the time dependence is induced by the coupling to the heat
bath. As may be seen from (\ref{equationofmotion}), such a term will
contribute to ${\cal{O}}(\lambda^2 g^2)$ to the evolution equation.
 We will consider {\it only} the first order  correction in $g^2$
 to the propagator arising from the influence functional, a typical diagram
 is depicted in figure 2(b).
 The one loop contributions to this order to
(\ref{equationofmotion}) are depicted in Figure 2(c).
Thus we will consistently neglect
to this order the second term on the right hand side of
 (\ref{massterm}) (time dependence of the order parameter).

The first one-loop contribution depicted in Figure 2(c) is the
familiar term, it is time independent and gives the usual result in
terms of the temperature of the field $\Phi$, for our purposes assumed to
be $T_{\Phi} \approx 0$. We
will absorb this term in a renormalization of the bare parameters,
$\mu^2 \; ; \; \lambda$.

Then we  define consistently to this order the time dependent mass
\ba
    m^2 (t) & = & (-\mu^2_R) + \Delta m(t) \label{massoft} \\
\Delta m(t) & = &  \frac{\lambda_R}{2} \int \frac{d^3
k}{(2\pi)^3}
 \langle  \varphi^+_{\vec{k}} (t) \varphi^+_{-\vec{k}}
(t) \rangle_{g^2} \label{deltamassoft}
\ea
where the ${\cal{O}}(g^2)$  contribution to the one-loop correlation
 function is depicted in figure 2(b). A simple analysis reveals, that
for $t \neq 0$, this one loop diagram is ultraviolet finite,
this contribution vanishes at $t=0$. Thus there are no further
renormalizations (up to this order) induced by the coupling to the heat
bath.

In the limit of $T_{\Phi} \approx 0$ we find
\be
  \Delta m(t) = \Delta m^{(0)}(t) +\Delta m^{(T_{\sigma})}(t)
\ee
where
\ba
  \Delta m^{(0)}(t)  & =      &   \frac{\lambda}{2} g^2 \int \frac{d^3 k}
{(2\pi)^3}
  \frac{1}{\omega^{\prime 2}_k \tilde{\omega}_k} \int^t_0
dt_1
 \int^{t_1}_0 dt_2
   \sin [ \omega'_k (t-t_1)] \left[
   \sin [ \tilde{\omega}_k (t_1-t_2)] \cos [\omega'_k (t-t_2)] \right.
   \nonumber\\
                     & +      & \left.  \cos
 [\tilde{\omega}_k (t_1-t_2)] \sin [\omega'_k
(t-t_2)] \right] \label{zerotemp} \\ [0.1in]
   m_{\Delta }^{(T \sigma)} (t)
                     &   =    &  \lambda g^2 \int \frac{d^3 k}
{(2\pi)^3}
  \frac{n_{\sigma}(\tilde{\omega}_k)}{\omega^{\prime 2}_k \tilde{\omega}_k}
 \int^t_0 dt_1 \int^{t_1}_0 dt_2 \left\{\sin [\omega'_k (t-t_1)]
 \cos[\tilde{\omega}_k (t_1-t_2)] \right. \nonumber \\
                     & \times & \left. \sin [\omega'_k (t-t_2)] \right\}
 \label{finitetemp} \\ [0.1in]
           \omega'_k & =      & \sqrt{\vec{k}^2+M^2} \; \; ; \; \;
    \tilde{\omega}_k  =  \sqrt{\vec{k}^2+\msigma^2} \label{frequencies}
\ea

Let us first consider the  finite temperature contribution. For this
purpose it proves convenient
 to introduce the following dimensionless variables,
\be
     x = \frac{k}{T_{\sigma}}  ~~~,~~~   \tau = T_{\sigma} t
\ee
After performing the time integrals in the high temperature limit
 $(T_{\sigma} \gg m_{\sigma} \; ; \;
M^2)$, the finite temperature contribution (\ref{finitetemp}) becomes

\be
  \Delta m^{(T_{\sigma})} (\tau) =\frac{\lambda T_{\sigma}^2}{2
\pi^2} \left(\frac{g^2}{\tsig^4}\right)
   \int^\infty_0 dx \frac{1}{x (e^x -1)} \left\{ \frac{1}{16
x^2}
 [ 1-\cos(2x\tau)]+ \frac{\tau^2}{8}[1-
\frac{\sin(2x\tau)}{x\tau}]
 \right\} \label{temptime}
\ee

This expression clearly shows that in this limit, the time scale for the
heat bath is $1/T_{\sigma}$.
As argued previously, we expect that for small system-bath coupling
($g/ \tsig^2 \ll 1$) the transition time $t_c  \gg 1/\tsig$.

For $\tau \gg 1$, the integrand in (\ref{temptime}) is sharply peaked at
$x \approx 0$ allowing a saddle-point approximation to the integral. We
find for $\tau \gg 1$

\be
  \Delta m^{(T_{\sigma})} (\tau) \approx   0.012 \tau^3 \lambda \tsig^2
 \left\{\frac{g^2}{\tsig^4}\right\} \label{onelooptemp}
\ee

In the same limit we estimate the
 contribution coming from the zero-temperature part to be
\be
  \Delta m^{(0)} (\tau)  \approx   0.013 \tau^2  \lambda \tsig^2
  \left\{\frac{g^2}{\tsig^4}\right\} \label{oneloopzero}
\ee

Thus as expected, in the $\tau \gg 1$ limit, the finite temperature
contribution  dominates.

It is  straightforward to see that in this limit the system-bath
dimensionless coupling is $g^2/ \tsig^4$, and that the terms that
we neglected in a perturbative expansion are of the order ${\cal{O}}
(\lambda^2\; , \;  \lambda^2(g^2/ \tsig^4) \; , \; \cdots )$.
 Recalling that the
critical temperature for the scalar field $\Phi$ is given by
$T_c^2 = 24 \mu_R^2 / \lambda_R$, it is convenient to rescale the
 field and
time to cast the equation in terms of dimensionless variables.
Define
\be
    s   =  \mu_R t \; \; ; \; \;
\chi^2  =  \frac{\lambda_R \phi^2}{6\mu_R^2}
\ee
the equation of motion (\ref{equationofmotion}) becomes
\ba
& & \frac{d^2 \chi(s)}{ds^2}-\chi(s)
\left[1-0.29s^3 \left(\frac{\tsig}{T_c}\right)^2 \left(\frac{g^2}
{\tsig\mu_R^3}\right)-0.31s^2\left(\frac{\tsig}{T_c}\right)^2
\left(\frac{g^2}{\tsig^2\mu_R^2}\right)\right]
\nonumber \\
& & +\chi^3 (s) - \frac{g^2}{\msigma\mu_R^3}\int_0^s ds'
 \sin\left[\frac{\msigma}{\mu_R}(s-s')\right] \chi(s')
 = 0  \label{dimensionlessequation}
\ea

Admittedly inserting the result from the one-loop term in the equation
of motion is not quite consistent at early times because the one-loop
results (\ref{onelooptemp},\ref{oneloopzero}) were obtained for large
$\tau = s(\tsig/\mu_R)$.  However, by integrating the above
equation we will obtain a qualitative understanding of the time evolution
of the order parameter.

Roughly speaking, the phase transition from
the broken symmetry state to the disordered phase takes place when the
effective time dependent mass (\ref{massoft}) becomes zero. This would
certainly be the case in the absence of the non-Markovian term (retarded
kernel in (\ref{equationofmotion})). The effect of this term may be
understood by expanding the field around the broken symmetry values
$\phi_{\pm} = \pm \sqrt{6\mu_R^2/ \lambda_R}$, and considering the deviation
from these values as perturbations. One finds that the retarded kernel
plays the role of a forcing term in the equation of motion for the
deviation, that tends to shift the value of the minima further away from
the origin.

However, we still expect that the phase transition will occur at time
scales when the effective mass term vanishes, as the oscillations of the
field will now occur near zero.

{}From the above effective equation of motion  we find that the effective
  mass
(\ref{massoft}) vanishes at the ``critical time'' $t_c$ given by
\be
t_c \approx \frac{1}{\tsig}\left[3.5\left(\frac{\tsig^4}{g^2}\right)
\frac{T^2_c}{\tsig^2}\right]^{\frac{1}{3}} \label{criticaltime}
\ee

This is one of the main results of this work. Although the time
scale is determined by the temperature of the heat bath, the critical
time is a non-analytic function of the coupling between the system and
bath.

Figure(3) shows the evolution of the equation of motion
 (\ref{dimensionlessequation}) for the  (arbitrary) value of the parameters
$\mu_R = \msigma \; ; \; g = \msigma^2 \; ; \;
 \tsig=T_c= 10 \msigma$. For these values, the critical time predicted
by (\ref{criticaltime}) is $\mu_R t_c \approx 5.8$, quite consistent with
the time at which the field crosses zero in Figure (3).

Thus we see that for $\tsig \approx T_c$ and $g^2/ \tsig^4 \ll 1$,
 $t_c \gg 1/\tsig$ (or alternatively $\tau_c \gg 1$)
 quite consistent with our original assumptions.

Notice that the critical time is non-perturbative in terms of the
system-bath coupling, in a sense very much like the critical temperature
is non-perturbative in terms of the scalar field self-coupling.
One would then argue that
the higher order corrections will
only contribute {\it perturbatively} to the critical time. This argument
however ignores the possible infrared divergences arising near the
 phase transition
as it happens in equilibrium finite temperature field theory
near the critical  point. So we expect that, very much as it happens in
equilibrium finite temperature, approaching the critical point or ``critical
time'' in a reliable manner will
 require a non-perturbative resummation of the infrared sensitive
diagrams\cite{weinberg,dolanjackiw,kapusta}.
Thus just as the critical temperature signals the breakdown of perturbation
theory to study the phase transition, this critical time may also signal
 the breakdown of perturbation theory in the system-bath coupling.
Presumably a non-perturbative resummation technique would have to be
invoked to study reliably the system-bath dynamics for times close to
the critical time $t_c$.

  This possibility
would have to be studied further and is beyond the scope of this article.

\section{\bf Semiclassical Langevin Equation}

A phenomenological but rather successful approach to study the
non-equilibrium dynamics of a particle coupled to a heat bath
 in
classical statistical mechanics is provided by the Langevin equation.
This approach provides a satisfactory description of Brownian motion of
a particle. The Langevin equation is the classical equation of
motion modified phenomenologically by (basically two) terms that account
 for the interaction with the heat bath. A term proportional to the
velocity of the particle that incorporates ``friction'' and dissipation,
and a stochastic force term that reflects the random ``kicks'' of the
heat bath upon the Brownian particle. In most applications,
 this stochastic
noise is assumed to be ``white'', that is completelety uncorrelated.
 The coefficient in the friction term
determines the relaxation time of the Brownian particle.
 The friction term and the stochastic force
are ultimately responsible for the approach to equilibrium of the particle
with the heat bath, and are thus related by the fluctuation-dissipation
theorem.

A very clear microscopic description leading to Langevin dynamics within
the context of one-particle quantum mechanics has
been presented by  Caldeira and Leggett and
 Schmid\cite{leggettcaldeira,schmidt}. Their study
reveals that in general the friction term arises as a local approximation
to a non-Markovian kernel for a particular choice for the density
of states of the heat-bath, and that the noise term is typically correlated
over time scales determined by the typical scales of the bath.

An attempt to obtain a microscopic description of Langevin dynamics
in {\it field theory} has been reported by Morikawa\cite{morikawa}
who considered a ``fermionic bath''.

In this section we offer a {\it semiclassical} derivation of the Langevin
 equation for the case of a bosonic heat-bath, emphasizing its range of
validity, and the formal steps to go beyond the semiclassical Langevin
equation.

Phenomenological  Langevin equations
have been recently used
to study the dynamics of semiclassical configurations  in a
 presence of a heat bath\cite{gleiser1,bochkarev,gleiser2}.
 Typically these Langevin
descriptions {\it assume} a friction term and a ``white-noise''. We
will show in what follows, that these assumptions may not be justified
in many cases, and that
a physically correct Langevin description must necessarily
incorporate details of the particular ``heat-bath''.

We now consider the case in which the $\Phi$ field is initially at
zero temperature, and it is coupled at $t=0$ to the heat bath, again
modelled by  the free field $\sigma$. The initial density matrix is now

\be
     {\hat{\rho}}(0) =  (|0\rangle_{\Phi} \langle 0|_{\Phi})
\otimes
 e^{-\beta_{\sigma}  H_{\sigma}}
\ee

Evolving this density matrix in time as in the previous section
(details are provided in the appendices), and tracing over the
bath variables (thus obtaining the influence functional) we obtain
the following generating functional

\be
Z[0]  =   \int  D \Phi_1  \int {\cal{D}} \Phi^+ {\cal{D}}
\Phi^- \left\{ e^{i  \int^{\infty}_{-\infty} d^4 x
{\cal{L}}_{\Phi} [\Phi^+]-
 {\cal{L}}_{\Phi} [\Phi^-]} \right\} \times {\cal{F}}
[\Phi^+,\Phi^-] \label{langevpart}
\ee

with the boundary conditions  $\Phi^+(\vec{r}, t= \infty) =
\Phi^-(\vec{r}, t=\infty) =\Phi_1(\vec{r})$.

At this stage it proves convenient to introduce the ``center-of-mass'' and
``relative'' field coordinates
 $\Psi$ and $\eta$,
$  \Psi = \frac{1}{2} [\Phi^+ + \Phi^-] ; \eta = [\Phi^+ -
\Phi^-]$. These are recognized as the coordinates used
in the Wigner transform of a density matrix\cite{feynmanvernon}.
The boundary conditions on the fields now become $\Psi(\vec{r},t= \infty)
= \Phi_1 \; ; \; \eta (\vec{r}, t=\infty)=0$, the integral over $\Phi_1$,
just defines the path integral over all field configurations, and
finally the partition function (\ref{langevpart}) becomes

\begin{eqnarray}
      Z[0]   & = & \int {\cal{D}} \Psi {\cal{D}} \eta  e^{iS[\Psi,\eta]}
 \nonumber\\
S[\Psi,\eta] & = & \int^{\infty}_{-\infty}  d^4x_1
\eta(x_1) \left[(-\Box -( -\mu^2)) \Psi (x_1) - \frac{\lambda}{3!}
\Psi^3 (x_1) \right. \nonumber \\
             &   & \left. -2 g^2 \int^{\infty}_{-\infty} d^4x_2
\Theta (t_1 -t_2)
{\cal{K}}_I (\vec{r}_1,t_1;\vec{r}_2,t_2) \Psi (\vec{r}_2,t_2)\right]
- \frac{\lambda}{4!}  \int^{\infty}_{-\infty}
 d^4x_1  \Psi(x_1) \eta^3(x_1)  +    \nonumber \\
             &   &  i\frac{g^2}{2}
\int^{\infty}_{-\infty} d^4x_1  \int^{\infty}_{-\infty} d^4x_2
\eta(x_1)
 {\cal{K}}_R (\vec{r}_1,t_1; \vec{r}_2,t_2) \eta(x_2)
 \label{langevin}
\end{eqnarray}
where the kernels ${\cal{K}}_{R}, {\cal{K}}_I$ are spatially
translationally invariant and their spatial Fourier transforms are given
by (see appendix A).
\ba
 {\cal{K}}_R (\vec{k},t_1,t_2)  & =  &
 \frac{1}{2 \tilde{\omega}_k}
 \cos[\tilde{\omega}_k (t_1-t_2)] [1+2n_{\sigma}
({\tilde{\omega}}_k)] \Theta (t_1) \Theta (t_2) \nonumber \\
 {\cal{K}}_I (\vec{k},t_1,t_2)  & =  &
 - \frac{1}{2 \tilde{\omega}_k}
 \sin [\tilde{\omega}_k (t_1-t_2)]\Theta(t_1)\Theta(t_2)
\end{eqnarray}
with $\tilde{\omega}_k$ being the frequencies of the heat-bath.
These kernels arise from the two-point correlation function of the
heat-bath fields, the real ${\cal{K}}_R$ and imaginary part
 ${\cal{K}}_I$ are related by the fluctuation-dissipation theorem.

To make contact with Langevin dynamics, it becomes convenient to
cast the quadratic term for the $\eta$ field as resulting from a
Gaussian integral over a stochastic noise term with probability
distribution given by

\be
    {\cal{P}} [ \xi] = \exp \left[- \frac{1}{2g^2}
 \left\{ \int^{\infty}_{-\infty}
  d^4x_1  \int^{\infty}_{-\infty}  d^4x_2 \xi (x_1) {\cal{K}}^{-1}_R
(x_1,x_2) \xi (x_2) \right\}\right]
\ee

Now the final form for the partition function becomes

\be
 Z[0]   =  \int {\cal{D}} \xi
 \int {\cal{D}} \Psi {\cal{D}} \eta {\cal{P}}[ \xi]
  e^{iS_{eff}[\Psi , \eta , \xi]}
\ee
with the effective action $S_{eff}$ given by
\begin{eqnarray}
  S_{eff}[\Psi  , \eta , \xi] & = & \int  d^4x_1  \eta(x_1)\left[(-\Box
-( -\mu^2))
 \Psi (x_1) - \frac{\lambda}{3!} \Psi^3 (x_1) \right. \nonumber \\
                              &   & \left. -2g^2 \int d^4x_2
\Theta (t_1 -t_2) {\cal{K}}_I (\vec{r}_1,t_1; \vec{r}_2,t_2) \Psi (x_2)
+ \xi (x_1) \right] \nonumber \\
                              & - &   \frac{\lambda}{4!}
  \int d^4x  \Psi(x) \eta^3(x) \label{langevinaction}
\end{eqnarray}

A semiclassical approximation to the above partition function requires
 the configurations that extremize the effective action. In particular
the condition $\delta S_{eff} / \delta \eta =0$ leads to the
{\it  lowest order} semiclassical result

\ba
 (\Box +( -\mu^2) )\Psi ({\vec{r}}_1,t_1) &+&
\frac{\lambda}{3!} \Psi^3
 ({\vec{r}}_1,t_1) + \nonumber \\
                                          & & 2 g^2
  \int dt_2 d^3 r_2 \Theta (t_1 -t_2)
 {\cal{K}}_I ({\vec{r}}_1,t_1;{\vec{r}}_2,t_2) \Psi
({\vec{r}}_2,t_2)
 = \xi({\vec{r}}_1,t_1) \label{langevinequation}
\ea
with the Gaussian noise correlation function
\be
\langle \xi ({\vec{r}}_1,t_1)
 \xi ({\vec{r}}_2,t_2) \rangle = g^2 {\cal{K}}_R
({\vec{r}}_1,t_1;{\vec{r}}_2,t_2) \label{noisecorrelation}
\ee

This is the typical Langevin equation. This equation, however, is not
the only dynamical evolution equation in the semiclassical limit, other
equations
are obtained by performing the variational derivatives with respect to the
``noise'' term and $\Psi$.  Langevin dynamics results from the decoupling
approximation to these equations, in particular neglecting the coupling
between $\Psi$ and $\eta$ and $\eta$ and the ``noise''. A consistent
improvement over the semiclassical Langevin equation will involve a
perturbative expansion in these terms. Allowing for the couplings of
 the different fields to the
``noise'' term will introduce corrections to the ``noise'' correlation
functions. This corrections are a manifestation of the ``back-reaction''
of the system on the bath correlations.

One may obtain an equation of motion similar to the effective equation
found in the previous section for the order parameter
 (\ref{equationofmotion}) from the above Langevin equation by splitting
the field as

\be
\Psi(\vec{r},t) = \phi(t) + \varphi(\vec{r},t)
\ee
requiring that the ``mean field'' $\phi(t)$ obeys a source free
equation and
considering $\varphi$ as the fluctuation whose equation of motion
 contains the noise term. The linearized equation for the fluctuation
with the noise term, and non-Markovian kernel may be solved by
 introducing the retarded propagator.
In the equation for the ``mean-field'' there is a term of the form

\[ \frac{\lambda}{2}\phi(t)(\varphi(\vec{r}_1,t))^2 \]

By taking the average of this term over the noise with the noise
auto-correlation
function (\ref{noisecorrelation}) one obtains the ${\cal{O}}(\lambda g^2)$
correction to the equation of motion of the order parameter,
 but {\it not} the one-loop correction without the
bath (${\cal{O}}(\lambda)$). This semiclassical Langevin equation does not
incorporate quantum loop corrections of the scalar self-interaction.

There are two features of the semiclassical Langevin equation that
deserve comment: the first is that the non-Markovian kernel does
not lead to dissipation. Secondly, the
noise does not have  ``white'' (delta functions) correlations. In fact
these two features are related by the fluctuation-dissipation theorem.

The fact that there is no dissipation is a consequence of the
simplicity of the model, one could try a coupling to the heat-bath
that is linear in the scalar field and quadratic in terms of the
bath fields. Integration over the bath fields will also yield to a
non-local influence functional which however will not be quadratic
in the scalar field, but may be studied perturbatively. The quadratic
contribution of the scalar field will involve a ``bubble'' diagram
from the bath fields. This diagram will have a two-particle threshold
that may lead to dissipative processes,
which however, is very high in energy if the bath particles are much
heavier than the system particles. Thus in this case long wavelength
low frequency components of the field will evolve without dissipation.

 Such a diagram appears naturally in the
case in which the bath field corresponds to fermions coupled to
the scalar field via a Yukawa coupling.
A  memoriless (Markovian) friction term in the Langevin equation
 (proportional to the time derivative of the field) may be obtained at
low frequencies provided the non-local kernel has a power series
expansion in the transferred frequency with a linear term in the
frequency. The coefficient of this linear term will be the friction
coefficient. From the above discussion, it is clear that
 this possibility is not very
easy to achieve if the bath particles are  more massive
than the system particles. In this case the multiparticle
 threshold will be very
high in frequency and there will be no dissipation for low transferred
frequency. This may be inferred from the work of
 Morikawa\cite{morikawa}. This situation may change if there are
collective excitations in the bath, that produce a linear frequency
dependence of the non-local kernel at low frequencies. This scenario,
however, is not generic and will depend on the details of the heat
bath.

This conclusion is particularly meaningful within the context of
Langevin dynamics of semiclassical field configurations. These
are large amplitude coherent configurations, but typically slowly
varying in space and time, thus corresponding to small frequency and
momentum transfers to the bath. Although very massive, because they are
mainly composed of long-wavelength, low frequency modes, it is difficult
for these configurations to ``decay'' and thus dissipate.

Derivative couplings between system and bath
fields may lead to dissipative terms, but are very dangerous from the
point of view of renormalizability. Couplings that are quadratic
(or higher order polynomials) in the
scalar field will yield to higher order polynomials in the influence
functional and certainly will not lead to a simple Langevin description.

Thus in order to obtain a semiclassical Langevin description there
is a strong restriction on the system-bath coupling: the system must
couple {\it linearly} to the bath, so that the influence functional is
quadratic. A friction term may arise if the bath degrees of
freedom are lighter than the system's degrees of freedom, but
as discussed by Morikawa\cite{morikawa} this will have to be
studied case by case, in particular it may require a higher order
calculation.

The point of this discussion is  that writing down a simple
Langevin equation with a local dissipative term and a noise term with
white correlations may have to be justified by looking at the particular
models in detail.

\section{\bf Conclusions:}

We have studied the dynamics of a phase transition in which a
scalar field theory (the system) , initially in the ordered
 (broken symmetry)
phase at low temperatures is coupled to a heat bath. This heat-bath
is represented
by another scalar field, linearly coupled to the ``system field''
and  whose temperature is close to the critical
temperature for the system.  We derived, the one-loop effective
 equations of motion
for the order parameter that incorporates the effects of the
heat bath.

 The heat bath introduces a non-Markovian, non-dissipative
 kernel that is temperature independent and also introduces one-loop
corrections that are temperature dependent.

The effective evolution equations were integrated and an estimate of the
time to complete the phase transition to the symmetric phase was
obtained. This time is a non-analytic function of the ``system-bath''
coupling and  may signal the breakdown of perturbation theory in terms of
this coupling.

We derived a semiclassical Langevin equation for the system, it
contains a non-Markovian, non-dissipative kernel, and the noise
term, although Gaussian (in this approximation), is colored and correlated
on time scales of the order of the inverse temperature of the heat-bath,
when this temperature is much larger than the masses of the fields.
It is pointed out that the non-Markovian kernel will be model dependent
and may only be approximated by a local dissipative (friction) term
only in very special cases. The derivation of the Langevin equation
permits to identify a formal expansion to improve over the semiclassical
approximation and to account for quantum effects (loops) of the system
field, as well as for back reaction of the system on the bath.

\vspace{2mm}

{\bf Acknowledgments:} The authors would like to thank D. Jasnow,
 R. Willey, A. Bochkarev, R. Carlitz, R. Holman, A. Weldon and J. Kapusta
for very enlightening conversations and comments. They also thank the
N.S.F for partial support through Grant No: PHY-8921311. D.-S. Lee
acknowledges partial support through a Mellon PreDoctoral Fellowship
Award.

\newpage

{\bf Appendix A : Finite Temperature Generating Functionals}

In this appendix, we summarize  the most relevant technical details
 leading to the generating functional (\ref{generatingfunction})
 in section 2. Most of the  steps may be found in the
literature\cite{jordan,niemisemenoff,kobeskowalski,calzetta,calzettahu},
 however we present here the generalizations and modifications
 appropriate to our case  in order to make our article self-contained.

Starting  from the generating functional (\ref{partfunc}) let us
insert the
resolution of the identity in terms of a
 complete set of field eigenstates
\be
 \int  D\Phi D\sigma  |\Phi,\sigma \rangle  \langle \sigma
,\Phi| = 1
\ee
between all the time evolution operators, obtaining

\begin{eqnarray}
Z[J^+_{\Phi}, J^-_{\Phi}, J^{\beta}_{\Phi}] & = & \int D \Phi_1 D
 \Phi_2 D \Phi_3
\int D \sigma_1 D \sigma_2  D \sigma_3 \nonumber\\
&  & \left\{ \langle \Phi_1,\sigma_1 |U_{\Phi}
 (T-i\beta_{\Phi},T;J^{\beta}_{\Phi}) U_{\sigma} (T-
i\beta_{\sigma},T)
 | \Phi_2,\sigma_2 \rangle \right\}\times \nonumber\\
&   & \left\{ \langle \Phi_2,\sigma_2 |U(T,T';J^-_{\Phi})|
 \Phi_3,\sigma_3\rangle   \langle \Phi_3,\sigma_3|
 U(T',T;J^+_{\Phi}) | \Phi_1,\sigma_1 \rangle\right\}
\nonumber\\
 &   &
 \end{eqnarray}
Each matrix element in the above expression has a functional integral
 representation. Then, the generating function becomes (to avoid
cluttering of notation we suppress the space-time indices)

\begin{eqnarray}
Z[J^+_{\Phi}, J^-_{\Phi}, J^\beta_{\Phi}] & = &  \int  D \Phi_1 D
\Phi_2 D
 \Phi_3
                           \int {\cal{D}} \Phi^+ {\cal{D}}
\Phi^-
 {\cal{D}}
                                       \Phi^\beta
\nonumber\\
&    & \left\{ e^{ i  \int_T^{T'}  {\cal{L}}_{\Phi} [\Phi^+] +
J^+_{\Phi}
 \Phi^+ - {\cal{L}}_{\Phi} [\Phi^-] - J^-_{\Phi} \Phi^-
}\right\}                                          \left\{
e^{i \int_T^{T-i\beta_{\Phi}}  {\cal{L}}_{\Phi} [\Phi^\beta] +
J^\beta_{\Phi} \Phi^\beta }  \right\} \times {\cal{F}}
[\Phi^+,\Phi^-]
                \nonumber\\
&     &
\end{eqnarray}
with the boundary conditions $\Phi^+ (T) = \Phi^\beta (T-
i\beta_{\Phi} ) =
 \Phi_1
; \Phi^- (T) = \Phi^\beta (T) = \Phi_2 $  and $\Phi^+ (T') =
\Phi^- (T')
 = \Phi_3 $.
This can be regarded as a path integral along the contour
$(C)$ in the
 complex time plane  shown in Figure $(1)$ with
periodic boundary  conditions.

The non-local functional ${\cal{F}}$ is  the ``influence
 functional''\cite{feynmanvernon,leggettcaldeira}
obtained by tracing-out the bath degrees of freedom ($\sigma$)

\begin{eqnarray}
 {\cal{F}} [\Phi^+,\Phi^-]& = &  \int  D \sigma_1 D \sigma_2
D \sigma_3
                           \int {\cal{D}} \sigma^+ {\cal{D}}
\sigma^-
 {\cal{D}}
                                       \sigma^\beta
\nonumber\\
   &   &     \left\{  e^{i  \int_T^{T'}  {\cal{L}}_{\sigma}
[\sigma^+] +
 {\cal{L}}_I
[\sigma^+,\Phi^+ ] - {\cal{L}}_{\sigma} [\sigma^-] -
{\cal{L}}_I [\sigma^-, \Phi^- ]  } \right\}
         \left\{ e^{i \int_T^{T-i\beta_{\Phi}}
{\cal{L}}_{\sigma} [\sigma^\beta]  }  \right\}
\end{eqnarray}
with the boundary conditions $\sigma^+ (T) = \sigma^\beta
(T-i\beta_{\sigma} ) = \Phi_1
; \sigma^- (T) = \sigma^\beta (T) = \sigma_2 $ and $\sigma^+
(T') = \sigma^- (T') = \sigma_3 $. The normalization term
 ${\cal{F}}[0,0]$ will cancel against the denominator in the computation
of correlation functions.
The influence functional  can also  be recognized as a path
integral along the contour $(C)$  shown in Figure (1).

For the linear coupling between system and bath fields, the trace
(path integral) over the bath variables can be done at once leading
to the expression (\ref{influfunc}).

The propagators $D_{\sigma}^{ab}$ $(a,b = +,-)$ are constructed
from the functions $D_{\sigma}^{>}$, $D_{\sigma}^{<}$  which are the
homogeneous
 solutions of the quadratic form in ${\cal{L}}_{\sigma} $
obeying the
 Kubo-Martin-Schwinger boundary condition (periodicity in imaginary
time)

\be
D_{\sigma}^{>}(\vec{r}_1,t_1-i\beta_{\sigma};\vec{r}_2,t_2) =
D_{\sigma}^{<}(\vec{r}_1,t_1;\vec{r}_2,t_2)
\ee
The spatial Fourier transforms of these functions are given by

\begin{eqnarray}
iD_{\sigma}^{>}(\vec{k};t,t') & = & \frac{1}{2
\tilde{\omega}_{k}}
 \left\{  (1 -
 n_{\sigma} ( \tilde{\omega}_{k})) e^{-i \tilde{\omega}_{k}(t-
t')} +
 n_{\sigma} ( \tilde{\omega}_{k}) e^{i \tilde{\omega}_{k}(t-
t')} \right\}
 \nonumber\\
D_{\sigma}^{>}(\vec{k};t,t')  & = &
D_{\sigma}^{<}(\vec{k},t',t) \nonumber \\
n_{\sigma} ( \tilde{\omega}_{k})&  =& \frac{1}
{e^{\beta_{\sigma} \tilde{\omega}_{k}} -1} \; \; ; \; \;
\tilde{\omega}_{k} =
 \sqrt{ {\vec{k}}^2 + m_{\sigma}^2} \label{sigmacorrel}
\end{eqnarray}

The Green's functions $( D^{ab}_{\sigma}(t_1,t_2)) $ that
enter in the integral are now explicitly  given by

\begin{eqnarray}
D_{\sigma}^{++}(\vec{r}_1,t_1;\vec{r}_2,t_2)  & = &
D_{\sigma}^{>}(\vec{r}_1,t_1;\vec{r}_2,t_2)\Theta(t_1-t_2) +
D_{\sigma}^{<}(\vec{r}_1,t_1;\vec{r}_2,t_2)\Theta(t_2-t_1)
\label{timeordered} \nonumber\\
D_{\sigma}^{--}(\vec{r}_1,t_1;\vec{r}_2,t_2)  & = &
D_{\sigma}^{>}(\vec{r}_1,t_1;\vec{r}_2,t_2)\Theta(t_2-t_1) +
D_{\sigma}^{<}(\vec{r}_1,t_1;\vec{r}_2,t_2)\Theta(t_1-t_2)
\label{antitimeordered} \nonumber \\
D_{\sigma}^{+-}(\vec{r}_1,t_1;\vec{r}_2,t_2)  & = & -
D_{\sigma}^{<}(\vec{r}_1,t_1;\vec{r}_2,t_2)
\label{plusminus} \nonumber\\
D_{\sigma}^{-+}(\vec{r}_1,t_1;\vec{r}_2,t_2)  & = & -
D_{\sigma}^{>}(\vec{r}_1,t_1;\vec{r}_2,t_2) = -
D_{\sigma}^{<}(\vec{r}_1,t_1;\vec{r}_2,t_2) \label{sigmagreen}
\end{eqnarray}

 As usual, the path
 integral over the quadratic form may be evaluated and we
may then expand all
  interaction terms perturbatively for  weak couplings.

After some straightforward algebra we obtain the following partition
 function
\begin{eqnarray}
Z[J^+_{\Phi}, J^-_{\Phi}, J^\beta_{\Phi}] & = &  \exp \left\{ i
 \frac{\lambda}{4!} \int^{T'}_T dt  [ (-i \frac{\delta}
{\delta J^+_{\Phi}})^4
 - (i\frac{\delta}{\delta J^-_{\Phi}})^4 ] \right\}  \exp
\left\{ i
 \frac{\lambda}{4!} \int^{T-i\beta_{\Phi}}_T dt  (-i
\frac{\delta}
{\delta J^\beta_{\Phi}})^4  \right\} \nonumber\\
     &   & {\cal{F}}[ -i \frac{\delta}{\delta  J^+_{\Phi}}, i
 \frac{\delta}{\delta  J^-_{\Phi}}] \times   \exp \left\{ -
\frac{i}{2}
  \int_{C} dt_1 \int_{C}  dt_2  J_{\Phi}^{C} (t_1)
D^{C}_{\Phi}(t_1,t_2)
 J_{\Phi}^{C}(t_2) \right\} \nonumber\\
      &   &
\end{eqnarray}
where $J^C_{\Phi}$ stands for the source term on the contour
$C$.

In the limit $T' \rightarrow \infty$ and,  $T\rightarrow  -
\infty$ the
 contributions from
the terms in which one of the currents is $J^+_{\Phi}$ or $J^-
_{\Phi}$ and
the other is a $J^{\beta}_{\Phi}$ vanish when computing
correlation
functions in which the external legs are at finite {\it real
time}, as a consequence of the Riemann-Lebesgue lemma\cite{mills}. For
this {\it real time} correlation functions, there is no
contribution from the $J^{\beta}_{\Phi}$ terms that cancel
between
numerator and denominator. Then the calculation of finite {\it
real time}
correlation functions, the generating functional simplifies
to
\begin{eqnarray}
Z[J^+_{\Phi}, J^-_{\Phi}] & = &  \exp \left\{ i
\frac{\lambda}{4!}
\int^{\infty}_{-\infty} dt  [ (-i \frac{\delta}{\delta
J^+_{\Phi}})^4 -
(i\frac{\delta}{\delta J^-_{\Phi}})^4 ] \right\}  {\cal{F}}[ -
i
 \frac{\delta}{\delta J^+_{\Phi}}, i \frac{\delta}{\delta  J^-
_{\Phi}}]
 \times  \nonumber\\
        &   & \exp \left\{ - \frac{i} {2} \int^{\infty}_{-
\infty}  dt_1
 \int^{\infty}_{-\infty} dt_2
           J^a_{\Phi} (t_1) D^{ab}_{\Phi}(t_1,t_2)  J_{\Phi}^a
 ( t_2) \right\}
 \label{generatingfunction2}
\end{eqnarray}
with $a,b = +,-$.

The Green's functions $D^{ab}_{\Phi}$ are similar to
 (\ref{sigmacorrel},\ref{sigmagreen}) and obtained from them by
replacing $m_{\sigma} \; ; \; \beta_{\sigma}$ by the values for the
$\Phi$ field.

\vspace{2mm}

{\bf Appendix B: Evolution equations}

The most convenient method to obtain the evolution equations is the
tadpole method\cite{weinberg}. In the present non-equilibrium situation,
it is implemented in the following manner. First we recall that the
non-equilibrium generating function without sources is given by

\be
Z[0] = \int  D \Phi_1 D \Phi_2 D \Phi_3
        \int {\cal{D}} \Phi^+ {\cal{D}} \Phi^- {\cal{D}} \Phi^\beta
    ~~~  e^{i {\cal{S}}_{eff} [\Phi^+ , \Phi^- , \Phi^\beta ]}
\ee
with the boundary conditions $\Phi^+ (T) = \Phi^\beta (T-i\beta_{\Phi} ) =
 \Phi_1
; \Phi^- (T) = \Phi^\beta (T) = \Phi_2 $  and $\Phi^+ (T') = \Phi^- (T')
 = \Phi_3 $. After integrating out the degrees of freedom of field
 $\sigma$ (bath), $ {\cal{S}}_{eff}$ including the influence functional
 is  thus
\begin{eqnarray}
 {\cal{S}}_{eff} [\Phi^+,\Phi^-, \Phi^\beta]
 &=&  \int_T^{T'} dt
 \{ {\cal{L}}_{\Phi} [\Phi^+]  - {\cal{L}}_{\Phi} [\Phi^-]\} +
 \int_T^{T-i\beta_{\Phi}} dt ~ {\cal{L}}_{\Phi} [\Phi^\beta] \nonumber\\
 &-& \frac{1}{2} g^2  \int^{T'}_0 dt_1 \int^{T'}_0 dt_2
           \Phi^a(t_1) D^{ab}_{\sigma}(t_1,t_2)  \Phi^a(t_2)
\label{effectiveaction}
\end{eqnarray}

 We shift the  $\Phi^\pm$  fields by
\be
\Phi^\pm  (\vec{r},t) = \phi (t) + \hat{\Phi}^\pm  (\vec{r},t)
\ee
where $\phi$ is a background mean field. The tadpole method\cite{weinberg}
 requires that
\be
      \langle \hat{\Phi}^\pm  (\vec{r},t)  \rangle  =  0 \label{tadpole}
\ee

We also need to shift the field
 $\Phi^\beta$ by
\be
\Phi^\beta  (\vec{r},t) = \phi (T) + \hat{\Phi}^\beta  (\vec{r},t)
\ee
Prior to the time when the system is coupled to the bath,
 the system is  in  equilibrium and  $\phi$ is a constant in time. So for
 $t \le 0\; , \; \phi (t)  = \phi(0)$ which is the initial equilibrium
 value of the mean field.

Now we expand

\[ {\cal{L}}(\Phi) = {\cal{L}}(\phi)+\frac{\delta {\cal{L}}}{\delta
\Phi}\Phi +\frac{1}{2}\frac{\delta^2 {\cal{L}}}{\delta \Phi^2}\Phi^2 +
\frac{1}{3!}\frac{\delta^3 {\cal{L}}}{\delta \Phi^3}\Phi^3+
\cdots \]

\noindent for the ($ + \; ,\; - \; , \; \beta$) branches
and consider the {\it linear}, cubic and quartic terms in $\Phi$ as
perturbations. The sources are now coupled to the fluctuation part of the
field to generate the perturbative expansion and require that
 order by order in perturbation theory
the tadpole condition (\ref{tadpole}) is fulfilled.

An important ingredient is the identity

\[ D^{++}+D^{+-}+D^{-+}+D^{--} =0 \]

for both the $\sigma$ and $\Phi$ Green's functions.

\newpage

\underline{\bf Figure Captions:}

\vspace{2mm}

\underline{\bf Figure 1:} Contour in the complex time plane for the
 non-equilibrium generating functional.

\vspace{2mm}

\underline{\bf Figure 2:} a) Propagators ($++$) for the ``system'' field
 (straight line) and ``bath'' field (wavy line), there are four of each
($++ \; ; +- \; ; -+ \; ; --$).
b) One loop correction to ${\cal{O}}(\lambda g^2)$. There are four terms
for the insertion of the bath propagators.

\vspace{2mm}

\underline{\bf Figure 3:} Numerical evolution of the effective equation
 of motion for $\chi(s)$ vs. $s$ for the values
$\mu_R= m_{\sigma} \; ; \; g = m^2_{\sigma} \; ; \; T_{\sigma}=T_c=10
m_{\sigma}$. The initial conditions are: $\chi(0)=1 \; ; \; \dot{\chi}(0)
=0$. The solid line is the evolution of the classical field equations
(without the non-Markovian kernel). The long dashed line is the evolution
including the non-Markovian kernel but without the one-loop correction.
The short dashed line is the full equation including the non-Markovian
kernel and the one-loop
contribution.


\begin{thebibliography}{99}
\bibitem{kolb}  E. W. Kolb and M. S. Turner, ``The Early Universe'', Addison
Wesley (Frontiers in Physics) (1990)
\bibitem{linde} See for example A. Linde, Particle Physics
and Inflationary Cosmology, Harwood Academic Publishers (1990),
and references therein.
\bibitem{brandenberger} R. H. Brandenberger, Rev. of Mod.
Phys. 57, 1 (1985); Int. J. Mod. Phys. 2A, 77 (1987).
\bibitem{starobinsky} A. A. Starobinsky, in Field Theory, Quantum
Gravity and Strings, Proceedings of the Seminar Series, Meudon and
Paris, France, 1984, Eds. H. J. de Vega and N. Sanchez, Lecture Notes
in Physics Vol. 246 (Springer, Berlin 1986).
\bibitem{paz1} E. Calzetta and B. L. Hu, Phys. Rev. D 35, 495 (1987);
37, 2838 (1988);
J. P. Paz, ibid. 41, 1054 (1990).
\bibitem{paz2} B. L. Hu in Quantum Mechanics in Curved Spacetime, Ed.
by J. Audretsch and V. de Sabbata (Plenum, London 1990), and Proceedings
of the Second International Workshop on Thermal Fields and Their
Applications, Ed. by H. Ezawa et. al. (North Holland, 1991)
\bibitem{paz3} B. L. Hu, J. P. Paz and Y. Zhang, Phys. Rev. D 45, 2843
(1992); ibid. 47, 1576 1993; T. A. Brun, ``Semiclassical equations of
motion for nonlinear brownian systems'', CALT-68-1848 (1993)
(unpublished).
\bibitem{zurek} W. H. Zurek, in Frontiers in Nonequilibrium Statistical
Physics, Ed. G. T. Moore and M. O. Scully (Plenum, N.Y. 1986); Phys. Rev.
D 24, 1516 (1981); 26, 1862 (1982); W. G. Unruh and W. H. Zurek, Phys.
Rev. D 40, 1071 (1989). G. W. Ford, J. T. Lewis and R.F. O'Connell,
Phys. Rev. A 37, 4419 (1988).
\bibitem{carlitz} R. D. Carlitz and R. Chakrabarti, Phys. Rev. D 31, 1418,
(1985).
\bibitem{cloutier} J. Cloutier and G. W. Semenoff, Phys. Rev. D 44,
3218 (1991). These authors consider symmetry restoration by coupling
linearly to a {\it noise} rather than another field that plays the
role of a heat bath.
\bibitem{boyveg} D. Boyanovsky and H. J. de Vega ``Quantum rolling down
out of equilibrium'', (to appear in Phys. Rev. D)
\bibitem{boyshin} D. Boyanovsky, D.-S. Lee and A. Singh
 ``Phase Transitions out of equilibrium: domain formation and growth''
(submitted); D. Boyanovsky ``Quantum spinodal decomposition''
 (submitted).
\bibitem{gleiser1} M. Alford, H. Feldman and M. Gleiser, ``Testing
classical nucleation theory'' (NSF-ITP-1992) (unpublished).
\bibitem{bochkarev} A. Bochkarev and P. de Forcrand, ``Non-perturbative
evaluation of the diffusion rate in field theory at high temperatures''
P. R. D. (to appear, 1993); Phys.Rev. D 44, 519 (1991); Phys. Rev. Lett.
63, 2337 (1989).
\bibitem{gleiser2} M. Alford, H. Feldman, and M. Gleiser, Phys. Rev.
Lett. 68, 1645 (1992).
\bibitem{leggettcaldeira} A. O. Caldeira and A. J. Leggett, Physica A
121, 587 (1983).
\bibitem{pazhu} J. P. Paz in Proceedings of the Second International
Workshop on Thermal Fields and Their Applications. Ed. H. Ezawa et. al.
(North Holland, 1991).
\bibitem{feynmanvernon} R. P. Feynman and F. L. Vernon, Jr. Ann. Phys.
(N.Y.) 24, 118 (1963).
\bibitem{grab} H. Grabert, P. Schramm and G. L. Ingold, Phys. Rep. 168,
115 (1988).
\bibitem{mills} R. Mills, ``Propagators for Many Particle
Systems''
(Gordon and Breach, N. Y. 1969).
\bibitem{schwinger} J. Schwinger, J. Math. Phys. 2, 407
(1961).
\bibitem{keldysh} L. V. Keldysh, Sov. Phys. JETP 20, 1018
(1965).
\bibitem{landsman} N. P. Landsman and C. G. van Weert, Phys.
Rep. 145, 141 (1987).
\bibitem{semenoffweiss} G. Semenoff and N. Weiss, Phys. Rev.
D31, 689; 699 (1985).
\bibitem{jordan} R. D. Jordan, Phys. Rev. D33, 444 (1986).
\bibitem{niemisemenoff} A. Niemi and G. Semenoff, Ann. of
Phys. (N.Y.) 152, 105 (1984); Nucl. Phys. B [FS10], 181, (1984).
\bibitem{kobeskowalski} R. L. Kobes and K. L. Kowalski,
Phys. Rev.
D34, 513 (1986); R. L. Kobes, G. W. Semenoff and N. Weiss,
Z. Phys. C 29, 371 (1985).
\bibitem{calzetta} E. Calzetta, Ann. of Phys. (N.Y.) 190, 32
(1989)
\bibitem{calzettahu} E. Calzetta and B. L. Hu, Phys. Rev.
D35, 495
(1987); Phys. Rev. D37, 2878 (1988).
\bibitem{weinberg} S. Weinberg, Phys. Rev. D9, 3357 (1974).
(1985).
\bibitem{dolanjackiw} L. Dolan and R. Jackiw, Phys. Rev. D 9, 3320
(1974)
\bibitem{kapusta} J. I. Kapusta, ``Finite Temperature Field Theory'',
Cambridge Univ. Press (1989).
\bibitem{schmidt} A. Schmid Jour. of Low Temp. Phys. 49, 609 (1982).
\bibitem{morikawa} M. Morikawa, Phys. Rev. D. 33, 3607 (1986).
\end{thebibliography}
\end{document}